\documentstyle[12pt]{article}

\begin{document}

\begin{flushright}
IMSc/2004/03/13 \\
hep-th/0404026
\end{flushright} 

\vspace{2mm}

\vspace{2ex}

\begin{center}
{\large \bf A Description of Schwarzschild Black Holes \\

\vspace{2ex}

in terms of Intersecting $M-$branes and antibranes } \\

\vspace{8ex}

{\large  S. Kalyana Rama}

\vspace{3ex}

Institute of Mathematical Sciences, C. I. T. Campus, 

Taramani, CHENNAI 600 113, India. 

\vspace{1ex}

email: krama@imsc.res.in \\ 

\end{center}

\vspace{6ex}

\centerline{ABSTRACT}
\begin{quote} 

Intersecting $M-$branes are known to describe multi charged
black holes. Using a configuration of such intersecting branes
and antibranes, together with massless excitations living on
them, we give a description of Schwarzschild black holes
following Danielsson, Guijosa, and Kruczenski. We calculate the
entropy of these black holes and find that it agrees, upto a
numerical factor, with the entropy of the corresponding
Schwarzschild black holes in supergravity approximation. We
give an empirical interpretation of this factor.

\end{quote}

\vspace{2ex}

%PACS numbers: 98.80.Cq

\newpage

\vspace{4ex}

{\bf 1.}  
In the last decade there has been a tremendous progress in
understanding the microscopic origin of the entropy and Hawking
radiation of extremal and near extremal black holes in terms of
various brane configurations in string/M theories and the low
energy excitations living on them \cite{bhreview}. In all these
cases, the regime of validity has been such that the
corresponding black holes have positive heat capacity. In
particular, the dynamics of Schwarzschild black holes remains
unexplained, although there have been a variety of other
attempts to understand Schwarzschild black holes
\cite{schbh,schbhrecent}.

Recently Danielsson, Guijosa, and Kruczenski (DGK) have given a
description of certain Schwarzschild black holes and an
explanation of their negative heat capacity \cite{dgk}. DGK
consider $D3, M2$, and $M5$ branes, stacks of which, in
supergravity approximation, correspond to single charged black
holes that become Schwarzschild ones when the net charge is
zero. DGK obtain these Schwarzschild black holes as a stack of
branes and an identical stack of antibranes, together with
massless excitations living on each stack. Assuming that such a
configuration alone suffices to describe the dynamics, DGK
obtain the resulting entropy of the Schwarzschild black hole and
provide an explanation of its negative heat capacity.\footnote{
It is perhaps of interest to note that negative heat capacities
have been observed experimentally in finite systems such as
atomic clusters and nuclear fragmentations \cite{sodium}.} The
agreement between this entropy and that of the Schwarzschild
black hole in supergravity approximation is almost complete:
these entropies differ by a numerical factor which is certain
power of $2$, numerically of ${\cal O}(1)$, and can be
interpreted to mean that the gas of massless excitations carried
twice the available energy \cite{dgk}.

It is natural to enquire whether a similar analysis can be
repeated for cases involving more than one type of branes, stacks
of which, in supergravity approximation, correspond to multi
charged black holes that become Schwarzschild ones when the net
charges are all zero. The extremal and near extremal limits of
such multi charged black holes are well described in string/M
theories using intersecting branes of different types
\cite{bhreview}.

In this letter, we study the case of intersecting $M-$brane
configurations which correspond to multi charged black
holes\footnote{ While our work was in final stages, there
appeared two papers \cite{peet,bl} which describe Schwarzschild
black holes a la DGK, using D-brane configurations corresponding
to single charged black holes.} \cite{ct,kt}. See also
\cite{ib}.  Following DGK, we obtain the corresponding
Schwarzschild black holes as stacks of intersecting $M-$branes
and an identical stack of intersecting antibranes, together with
massless excitations living on each stack. Assuming that such a
configuration alone suffices to describe the dynamics, we obtain
the resulting entropy of the Schwarzschild black hole. Its
negative heat capacity has a similar origin as in \cite{dgk}. We
find that this entropy agrees with that of the Schwarzschild
black hole in supergravity approximation, upto a numerical
factor which is certain power of $2$, is numerically of ${\cal
O}(1)$, and generalises that in \cite{dgk}. We give an empirical
interpretation of this factor which differs from that in
\cite{dgk}.

This letter is organised as follows. In sections {\bf 2} and
{\bf 3}, we give a brief description of the relevent results of
\cite{dgk} and of the multi charged black holes described by
intersecting $M-$branes, following \cite{ct,kt}. In section {\bf
4}, we obtain the entropy of the corresponding Schwarzschild
black holes. In section {\bf 5}, we give an empirical
interpretation of the factor mentioned above. In section {\bf
6}, we conclude by mentioning a few issues for further study.

{\bf 2.}  
We give a brief description of the relevent results of
\cite{dgk}. DGK consider stacks of $p-$branes which, in
supergravity approximation, correspond to single charged black
holes in transverse $D-$dimensional spacetime. They obtain the
corresponding $D-$dimensional Schwarzschild black holes as a
system consisting of (i) a stack of branes $N$ in number; (ii)
an identical stack of antibranes also $N$ in number; and (iii)
two copies, one on each stack, of a gas of massless excitations
$\propto N^c$ in number. The two copies of the gas are assumed
not to interact with each other. DGK assume that such a system
alone suffices to describe the dynamics of Schwarzschild black
holes. For the $D3, M2$, and $M5$ branes considered in
\cite{dgk} \begin{equation}\label{dpc} (D; p, c) = (7; 3, 2) \;
, (9; 2, \frac{3}{2}) \; , (6; 5, 3) \end{equation}
respectively. For a complete discussion and further details see
\cite{dgk}.

The branes, and similary antibranes, have zero entropy and
energy $= C N$ where the constant $C$ includes tension and
volume of the branes. Each copy of the gas has energy $E$,
entropy $S$, and temperature $T$ which obey the following
equivalent relations:
\begin{eqnarray}
S & = & B N^c T^p \; , \; \; \; 
E = \frac{p}{p + 1} B N^c T^{p + 1} \label{set} \\
S & = & A N^\alpha E^a  \; , \; \; \; 
\frac{1}{T} = \frac{a S}{E} 
\; \; \; {\rm where} \; \; \; 
\alpha = \frac{c}{p + 1} \; , \; \; \; 
a = \frac{p}{p + 1} \; , \label{se}
\end{eqnarray} 
$A$ can be obtained easily in terms of $B$ and $p$, and $B$
depends on tension, volume, etc. and is given in \cite{dgk}.
The total energy $M$ and the total entropy $S_{tot}$ of the
system are then given by
\begin{equation}\label{stot}
M = 2 C N + 2 E \; , \; \; \; S_{tot} = 2 S(E) 
= 2^{1 - a} A N^\alpha (M - 2 C N)^a \; .
\end{equation}

In canonical formalism, such a system is unstable towards
creating an infinite number of brane antibrane pairs. Hence, one
must work in microcanonical formalism where the total energy $M$
of the system is kept fixed and the equilibrium quantities are
obtained by maximising the entropy $S_{tot}$ of the system with
respect to $N$. This then determines $N$ and $E$ to be 
\begin{equation}\label{ne0}
2 C N = \frac{\alpha M}{a + \alpha} \; , \; \; \; 
2 E = \frac{a}{\alpha} \; (2 C N) = 
\frac{a M}{a + \alpha} \; . 
\end{equation} 
Substituting these equilibrium values in the expression for
$S_{tot}$ given in equation (\ref{stot}) one finds that
\begin{equation}\label{smax}
S_{tot}(M) = 2^{1 - a - \alpha} a^a \alpha^\alpha C^{- \alpha} \; 
A \left( \frac{M}{a + \alpha} \right)^{a + \alpha} \; . 
\end{equation} 
The temperature $T_{tot}$ of the system is given by 
\[
\frac{1}{T_{tot}} = \frac{(a + \alpha) S_{tot}}{M}
= \frac{a S}{E} = \frac{1}{T} 
\]
and is the same as the temperature $T$ of the gas given in
equation (\ref{se}). 

In equation (\ref{smax}) the exponent of $M$ is given by $(a +
\alpha) = \frac{c + p}{p + 1}$. Therefore, the heat capacity of
the system is negative for the $D3, M2$, and $M5$ branes
considered in \cite{dgk} since $c$ and, hence, $a + \alpha > 1$,
see equation (\ref{dpc}). Indeed, for these cases
\begin{equation}\label{dgk} 
a + \alpha = \frac{c + p}{p + 1} = \frac{5}{4} \; , 
\frac{7}{6} \; , \frac{4}{3} 
\end{equation}
respectively. The $M-$dependence of $S_{tot}$ is thus the same
as that of a $D-$dimensional Schwarzschild black hole with $D =
7, 9, 6$, given in equation (\ref{dpc}). Working out explicitly
the other constants, it turns out that the entropy $S_{tot}(M)$
in equation (\ref{stot}) and the entropy $S_0(M)$ of the
Schwarzschild black hole of the same mass $M$ in supergravity
approximation are equal upto a deficit factor $2^{- \frac{p}{p +
1}}$:
\begin{equation}\label{x0}
S_{tot}(M) = 2^{- \frac{p}{p + 1}} S_0(M) \; . 
\end{equation}
This deficit factor can be interpreted to mean that the gas of
massless excitations on the branes and antibranes has twice the
available energy \cite{dgk}.

DGK also provide an explanation of the negative heat capacity.
In equilibrium, the system maintains a fixed ratio between the
gas energy and brane antibrane energy, see equation (\ref{ne0}).
When the total energy $M$ of the system decreases, the gas
energy decreases, so must the brane energy. It is then
entropically favourable for a brane antibrane pair to
annihilate, giving energy to the gas thereby increasing its
temperature. It is essentially this process that results in
negative heat capacity \cite{dgk}. 

{\bf 3.}
DGK consider $D3, M2$, and $M5$ branes. In supergravity
approximation, stacks of these branes describe well the extremal
and near extremal limits of single charged black holes which
have zero entropy in the extremal limit. Stacks of such branes
and identical stacks of antibranes, together with massless
excitations living on each stack, then describe the
corresponding Schwarzschild black holes.

It is natural to enquire whether a similar analysis can be
repeated for cases involving more than one type of intersecting
branes, stacks of which, in supergravity approximation, describe
well the extremal and near extremal limits of multi charged
black holes which have zero entropy in the extremal
limit. Stacks of such intersecting branes and identical stacks
of intersecting antibranes, together with massless excitations
living on each stack, should then describe the corresponding
Schwarzschild black holes. This will then generalise the results
of \cite{dgk} to a wider class of black holes.

It turns out that such a description is indeed possible. In this
letter, we study intersecting $M-$brane configurations which, in
supergravity approximation, describe the extremal and near
extremal limits of multi charged black holes. We consider only
those cases where the entropy is zero in the extremal limit.

Following \cite{ct,kt}, consider intersecting M-brane
configurations in 11 dimensional spacetime with $p$ compact
directions and let $D = 11 - p$.  Various M-brane configurations
are given in \cite{ct,kt} that dimensionally reduce to black
holes with $K$ number of charges in $D$ dimensional spacetime,
$3 < D < 11$. For example, intersection of three stacks of M5
branes with $(D, K) = (4, 3)$; intersection of two stacks of M5
branes with $(D, K) = (4, 2)$; intersection of two stacks of M2
and M5 branes with $(D, K) = (5, 2)$; intersection of two stacks
of M2 branes with $(D, K) = (7, 2)$; etcetera. See \cite{ct,kt}
for a complete list. Our analysis and results below apply to all
the cases given in \cite{ct,kt} for which $\lambda = \frac{D -
2}{D - 3} - \frac{K}{2} > 0$.

Let $r$ be the radial coordinate in the transverse space. Then,
in Einstein frame, the $D-$dimensional transverse spacetime
metric is given by \cite{ct,kt}
\begin{eqnarray} 
d s_D^2 & = & h^{\frac{1}{D - 2}} \left( - \frac{f}{h} d t^2 
+ \frac{d r^2}{f} + r^2 d \Omega^2_{D - 2} \right) \nonumber \\
f & = & 1 - \frac{2 \mu}{r^{D - 3}} \; , \; \; \; 
h = \prod_{i = 1}^K \left( 1 + \frac{{\cal Q}_i}{r^{D - 3}} 
\right) \nonumber \\
{\cal Q}_i & = & \sqrt{Q_i^2 + \mu^2} - \mu \; , \; \; \; 
Q_i \equiv A_i N_i \label{2.1} 
\end{eqnarray} 
where $N_i$ is the number of $i^{th}$ type brane and $A_i$ are
constants. The mass and the entropy are given by
\begin{eqnarray}
M_{sg} & = & b \left(2 \lambda \mu + 
\sum_{i = 1}^K \sqrt{Q_i^2 + \mu^2} \right) \; , \; \; \; 
S_{sg} = \frac{4 \pi b (2 \mu)^\lambda}{D - 3} \; 
\prod_{i = 1}^K \sqrt{{\cal Q}_i + 2 \mu} \nonumber \\ 
b & = & \frac{(D - 3) \omega_{D - 2} V_p}{2 G_{11}} 
\; , \; \; \; \lambda = \frac{D - 2}{D - 3} - \frac{K}{2} \; . 
\label{2.2}
\end{eqnarray}
In the above equations, $G_{11}$ is the $11-$dimensional Newton's 
constant, $V_p$ is the volume of the $p-$dimensional compact space, 
and $d \Omega_{D - 2}$ and $\omega_{D - 2}$ are the line element 
and surface area of a unit $(D - 2)-$dimensional sphere. 

Schwarzschild solution is obtained by setting $Q_i = 0$. The mass 
and the entropy are then given by 
\begin{equation}\label{sch}
M_0 = \frac{b (D - 2)}{D - 3} \; 2 \mu \; , \; \; \; 
S_0 = \frac{4 \pi b}{D - 3} \; 
(2 \mu)^{\frac{D - 2}{D - 3}} \; . 
\end{equation} 
The extremal solution is obtained by setting $\mu = 0$. The mass and 
entropy are then given by 
\begin{equation}\label{ext}
M_e = b \sum_{i = 1}^K Q_i \; , \; \; \; 
S_e = 0 
\end{equation} 
since we take $\lambda > 0$ as we are interested only in those
cases where extremal entropy is zero.

By considering the near extremal limit where $\mu$ is small, the
dynamics of the intersecting brane configurations can be
obtained from the above solutions and can be thought of as
arising due to a gas of massless excitations. The precise nature
of excitations depends on the configuration. For example, for
intersecting $M5$ branes, they consist of particles in $(1 +
1)-$dimensions, or fluctuating strings, or fluctuating 3-branes
depending on the value of $K$. See \cite{kt} for details.
Defining the energy of the gas on the branes to be $E = M_{sg} -
M_e$, one obtains
\begin{equation}\label{ne}
E = M_{sg} - M_e = 2 \lambda b \mu \; , \; \; \; 
S = \frac{4 \pi b}{D - 3} \; \left( 
\prod_{i = 1}^K \sqrt{Q_i} \right) \; (2 \mu)^\lambda \; . 
\end{equation}
We write the above equations as 
\begin{equation}\label{sbeb}
S(E) = A \left( \prod_{i = 1}^K \sqrt{Q_i} \right) 
E^\lambda \; \; \; {\rm where} \; \; \; 
A \equiv \frac{4 \pi b (\lambda b)^{- \lambda}}{D - 3} \; . 
\end{equation}
Then the temperature of the gas is given by 
\begin{equation}\label{t}
\frac{1}{T} = \frac{\lambda S}{E}
\end{equation}
and the entropy of the Schwarzschild black hole in (\ref{sch})
can be written as
\begin{equation}\label{s0m0}
S_0(M_0) = A \lambda^\lambda b^{- \frac{K}{2}}
\left( \frac{M_0}{\lambda + \frac{K}{2}} 
\right)^{\lambda + \frac{K}{2}} 
= A \lambda^\lambda b^{- \frac{K}{2}}
\left( \frac{(D - 3) M_0}{D - 2} 
\right)^{\frac{D - 2}{D - 3}} \; . 
\end{equation}

{\bf 4.}
We now describe, following \cite{dgk}, the corresponding
$D-$dimensional Schwarzschild black holes as a system consisting
of (i) a stack of intersecting $M-$branes, $N_i$, $i = 1, 2,
\cdots, K$ in number; (ii) an identical stack of intersecting
antibranes, also $N_i$ in number; and (iii) two copies, one on
each stack, of a gas of massless excitations, whose entropy and
energy are related by equation (\ref{sbeb}). The two copies of
the gas are assumed not to interact with each other. We assume
that such a system alone suffices to describe the dynamics of
Schwarzschild black holes.

The intersecting branes, and similary intersecting antibranes,
have entropy $ = 0$ and energy $= b \sum_{i = 1}^K Q_i$. Each
copy of the gas has energy $E$, entropy $S$, and temperature $T$
which are related by equations (\ref{sbeb}) and (\ref{t}). The
total energy $M$ and the total entropy $S_{tot}$ of the present
system are given by
\begin{equation}\label{mstot}
M = 2 b \sum_{i = 1}^K Q_i + 2 E \; , \; \; \; 
S_{tot} = 2 S(E) = 2 A 
\left( \prod_{i = 1}^K \sqrt{Q_i} \right) E^\lambda 
\end{equation}
where $Q_i = A_i N_i$, $i = 1, \cdots, K$, with the constants
$b A_i$ including tension and volume of the branes. As in
\cite{dgk}, in canonical formalism, such a system is unstable
towards creating an infinite number of brane antibrane
pairs. Hence, one must work in microcanonical formalism where
the total energy $M$ of the system is kept fixed and the
equilibrium quantities are obtained by maximising the entropy
$S_{tot}$ of the system with respect to $N_i$, equivalently
$Q_i$, for $i = 1, \cdots, K$.  This then determines $N_i$ and
$E$ to be
\begin{equation}\label{nie}
2 E = \frac{2 \lambda M}{2 \lambda + K} \; , \; \; \; 
2 b Q_i = \frac{M}{2 \lambda + K} \; . 
\end{equation}
Substituting these equilibrium values in the expression for
$S_{tot}$ given in equation (\ref{mstot}) we find that
\begin{eqnarray}
S_{tot}(M) & = & 2^{1 - \lambda - K} 
A \lambda^\lambda b^{- \frac{K}{2}}
\left( \frac{M}{\lambda + \frac{K}{2}} 
\right)^{\lambda + \frac{K}{2}} \nonumber \\
& = & 2^{1 - \lambda - K} A \lambda^\lambda 
b^{- \frac{K}{2}} \left( \frac{(D - 3) M_0}{D - 2} 
\right)^{\frac{D - 2}{D - 3}} \; . \label{s}
\end{eqnarray}
The temperature $T_{tot}$ of the system is given by  
\[
\frac{1}{T_{tot}} = \frac{(2 \lambda + K) S_{tot}}{2 M}
= \frac{\lambda S}{E} = \frac{1}{T} 
\]
and is the same as the temperature $T$ of the gas given in
equation (\ref{t}).

In equation (\ref{s}) the exponent of $M$ is $> 1$. Therefore,
the heat capacity is negative for the system of intersecting
branes antibranes described above. The explanation of the
negative heat capacity is the same as that provided in
\cite{dgk} and which is given briefly in section {\bf 2}.

In supergravity approximation, the entropy $S_0(M)$ of the
Schwarzschild black hole of the intersecting brane antibrane
configurations with mass $M$ and with all charges set to zero is
obtained from equation (\ref{sch}) or, equivalently,
(\ref{s0m0}). Comparing it with the total entropy $S_{tot}(M)$
of the intersecting brane antibrane configurations of the same
mass $M$, obtained above in equation (\ref{s}), we see that
\begin{equation}\label{sfinal}
S_{tot}(M) = 2^{1 - \lambda - K} S_0(M) 
\equiv X S_0(M) \; . 
\end{equation} 
That is, these two entropies are equal upto a deficit factor $X
\equiv 2^{1 - \lambda - K}$. For $K = 1$ we have $X = 2^{-
\lambda}$ which is the factor found in \cite{dgk} since, for $D
= 9$ and $6$ we have $(p, \lambda) = (2, \frac{2}{3})$ and $(5,
\frac{5}{6})$ respectively, see equation (\ref{x0}). We give an
empirical interpretation of this deficit factor $X$ in the next
section.

{\bf 5.}
For $D3, M2$, and $M5$ brane antibrane configurations, the
deficit factor $X$ can be interpreted to mean that the gas on
the brane and antibranes carried twice the available energy
\cite{dgk}. For intersecting brane antibrane configurations,
such an interpretation is clearly not sufficient for $K > 1$. To
understand this deficit factor further, we study how the
resulting total entropy $S_{tot}$ changes if one normalises the
brane tension, the gas energy, and entropy by constant factors.
For this, we consider the total energy and entropy of the
configuration to be given by
\begin{equation}\label{mstotscaling} 
M = 2 \alpha b \sum_{i = 1}^K Q_i + 2 \gamma E \; , \; \; \; 
S_{tot} = 2 \sigma A \left( \prod_{i = 1}^K \sqrt{Q_i} \right) 
(\epsilon \gamma E)^\lambda \; . 
\end{equation}
The factor $\alpha$ normalises the brane tensions,\footnote{ In
equation (\ref{mstotscaling}), $\alpha$ normalises only the
total brane energy which includes tensions, volumes, and number
of branes. However, it is natural to take $\alpha$ as
normalising brane tensions, and thereby the total brane energy.}
$\gamma$ and $\epsilon$ the gas energies, and $\sigma$ the gas
entropy. Maximising as before the total entropy $S_{tot}$ with
respect to $N_i$, equivalently $Q_i$, determines $N_i$ and $E$
to be
\[
2 \gamma E = \frac{2 \lambda M}{2 \lambda + K} \; , \; \; \; 
2 \alpha  b Q_i = \frac{M}{2 \lambda + K} \; . 
\]
Substituting these equilibrium values in the expression for
$S_{tot}$ in equation (\ref{mstotscaling}), we find that 
\begin{equation}\label{sfinalscaling}
S_{tot}(M) = 2^{1 - \lambda - K} \sigma \epsilon^\lambda 
\alpha^{- \frac{K}{2}} \; S_0(M) 
\equiv X S_0(M) \; .
\end{equation}
Thus, if the deficit factor $X = 1$ then the total entropy
$S_{tot}(M)$ of the intersecting brane antibrane configurations
and the entropy $S_0(M)$ of the corresponding Schwarzschild
black hole in supergravity approximation will be exactly
equal. $X = 2^{1 - \lambda - K} \sigma \epsilon^\lambda
\alpha^{- \frac{K}{2}}$ will be $ = 1$ for any value of $K$ and
$D$, equivalently $\lambda$, if we set
\[ 
\epsilon = 2 \; , \; \; \;  \sigma = \frac{1}{2} 
\; , \; \; \; \alpha = \frac{1}{4} \; .  
\]
$\epsilon = 2$ and $\sigma = \frac{1}{2}$ means that the gas
energy is to be increased by a factor of $2$ and the gas entropy
decreased by a factor of two. Empirically, they can be taken
together to mean simply that the available gas energy is not
shared equally by the two copies of massless excitations on
branes and antibranes but, instead, is all taken by only one
copy - or, more generally, by one linear combination of the two
copies; the other copy is frozen out perhaps by becoming
massive. $\alpha = \frac{1}{4}$ means that the brane tension is
to be decreased by a factor of $4$, for which we are not able to
give a simple empirical interpretation.

Note that this interpretation of the deficit factor $X$ differs
from that of \cite{dgk}. For the cases considered by DGK, the
brane tensions appear with an exponent $= - \frac{c}{p + 1}$ in
the expression for the maximised total entropy $S_{tot}$, see
\cite{dgk}. It turns out that this exponent $= - \frac{1}{2}$
for all the cases considered there, see equation (\ref{dpc}).
Therefore, decreasing the tensions by a factor of $4$ has the
effect of increasing $S_{tot}$ by a factor of $2$, which
prescisely compensates the $\sigma = \frac{1}{2}$ factor. The
net effect for the cases in \cite{dgk} is then simply to
increase the gas energy by a factor of $2$.

Here we see that the entropy of the intersecting brane antibrane
configurations will be equal to that of the corresponding
Schwarzschild black holes of the same mass in supergravity
approximation if in the brane antibrane configurations (i) the
tensions of all the branes are decreased by a factor of $4$ and
(ii) of the two copies of the gas of massless excitations only
one copy - more generally, one linear combination - participates
in the dynamics, the other copy being frozen out perhaps by
becoming massive.

Optimistically speaking, it is conceivable that the above
interpretation somehow captures the essence of brane antibrane
dynamics relevent for a description of Schwarzschild black
holes. On the other hand, the deficit factor $X$ may simply be
due to the `binding energy' of branes and antibranes, an
interpretation advocated recently in \cite{peet}. To understand
this issue fully one needs a rigorous study of brane antibrane
dynamics at finite temperature which, however, is likely to
require the full arsenel of string field theory techniques
\cite{sftreview}.

{\bf 6.}
We find that a class of black holes can be described using
intersecting $M-$brane antibrane configurations, generalising
those in \cite{dgk}. The agreement between the entropy of the
intersecting brane antibrane configurations and that of the
corresponding Schwarzschild black hole in supergravity
approximation is impressive. But these two entropies differ by a
deficit factor. We provide an empirical interpretation of it,
which is different from that of \cite{dgk}.

We conclude by mentioning a few issues that can be studied
further. By a chain of $S, T, U$ dualities, intersecting
$M-$brane configurations can be related to a variety of brane
configurations in string theory \cite{ib}. In view of the
results here and in \cite{dgk,peet,bl}, and as our preliminary
calculations also indicate, it is quite likely that
corresponding Schwarzschild black holes can be described by
intersecting brane antibrane configurations. Some of them may
have non trivial dilaton, the role of which may perhaps be
better understood by relating back these configurations to the
$M-$brane antibrane configurations presented here.

More detailed description of Schwarzschild black holes requires
a better understanding of finite temperature $D-$ and $M-$brane
antibrane dynamics, where massless excitations may involve
fluctuating strings and fluctuating $3-$branes. String field
theory techniques \cite{sftreview} are essential towards a study
of such issues. For some ideas in this context, see
\cite{dgk,peet} and references therein.

%\vspace{3ex}

{\bf Acknowledgement:} 
We thank N. Ohta for pointing out some of the references 
in \cite{ib}.

%\newpage

\end{document}